\documentclass[runningheads]{llncs}

\usepackage{comment}
\usepackage{float}

\usepackage{amssymb}
\usepackage{amsmath}
\usepackage{graphicx}
\usepackage{xcolor}
\usepackage{url}
\usepackage{hyperref}
\usepackage{tabularx}
\usepackage{subcaption}
\usepackage{makecell}
\usepackage{cite}
\usepackage{listings}
\usepackage{esint}
\usepackage{array}
\usepackage{float}
\usepackage{multirow}
\usepackage{varwidth}
\usepackage[colorinlistoftodos]{todonotes}
\usepackage{enumitem}
\usepackage{caption}
\definecolor{mygreen}{RGB}{0,160,25}

\urldef{\mailsb}\path|jan.valdman@utia.cas.cz |

\providecommand{\keywords}[1]{\textbf{\textit{Keywords:}} #1}

\definecolor{myart}{RGB}{0,0,0}

\def\BibTeX{{\rm B\kern-.05em{\sc i\kern-.025em b}\kern-.08em
    T\kern-.1667em\lower.7ex\hbox{E}\kern-.125emX}}
    \usepackage{comment}

\newcommand\dx{\, \mathrm{d}x}

\providecommand{\tabularnewline}{\\}
\newenvironment{cellvarwidth}[1][t]
    {\begin{varwidth}[#1]{\linewidth}}
    {\end{varwidth}}

\definecolor{defretcolor}{rgb}{0.00, 0.42, 0.24} 
\definecolor{jnpcolor}{rgb}{0.86, 0.08, 0.24} 
\definecolor{funccolor}{rgb}{0.06, 0.25, 0.49} 
\definecolor{numbercolor}{rgb}{0.48, 0.00, 0.48} 
\definecolor{commentcolor}{rgb}{0.38, 0.63, 0.69} 
\definecolor{backcolour}{rgb}{0.95, 0.95, 0.96} 
\definecolor{stringcolor}{rgb}{0.31, 0.60, 0.02} 

\lstdefinestyle{mystyle}{
    backgroundcolor=\color{backcolour},
    commentstyle=\color{commentcolor},
    basicstyle=\ttfamily\small,
    breakatwhitespace=false,
    breaklines=true,
    captionpos=b,
    keepspaces=true,
    numbers=left,
    numbersep=5pt,
    showspaces=false,
    showstringspaces=false,
    showtabs=false,
    tabsize=2,
    keywordstyle=\color{black}, 
    stringstyle=\color{stringcolor},
    classoffset=0, 
    morekeywords={def, return}, 
    keywordstyle=\color{defretcolor},
    emph={array, sum, at, abs, log, set, dot}, 
    emphstyle=\color{funccolor},
    classoffset=0, 
    morekeywords={jnp},
    keywordstyle=\color{jnpcolor},
    classoffset=2, 
    morekeywords={0,1,2,3,4,5,6,7,8,9},
    keywordstyle=\color{numbercolor}, 
    classoffset=0, 
}

\title{Minimization of Nonlinear Energies in Python Using 
FEM and Automatic Differentiation Tools
\thanks{
J. Valdman was supported by the project grant
24-10366S (GA\v{C}R) on Coupled dissipative processes and deformation mechanisms in metastable titanium alloys.
}
}

 \titlerunning{Minimization of Nonlinear Energies in Python}

\author{Michal Béreš\inst{1,2}\orcidID{0000-0001-8588-3268}, Jan Valdman\inst{3,4}\orcidID{0000-0002-6081-5362}}
\authorrunning{Michal Béreš, Jan Valdman}
\institute{Institute of Geonics of the Czech Academy of Sciences,
Studentská~1768, 
Ostrava,
708~00, 
Czech Republic \\
\email{michal.beres@ugn.cas.cz}
\and
Department of Applied Mathematics, VSB - Technical University of~Ostrava,
17.~listopadu~15/2172, 
Ostrava,
708 00, 
Czech Republic
\and
Institute of Information Theory and Automation of the Czech Academy of Sciences, 
Pod vod\'{a}renskou v\v{e}\v{z}\'{\i}~4, 18208 Prague, Czech Republic \\
\email{jan.valdman@utia.cas.cz}
\and
Department of Computer Science, Faculty of Science, University of South Bohemia, 
Brani\v sovsk\' a 31, 37005~\v{C}esk\'{e}~Bud\v{e}jovice, Czech Republic
\\
}

\begin{document}

\maketitle

\begin{abstract}
This contribution examines the capabilities of the Python ecosystem to solve nonlinear energy minimization problems, with a particular focus on transitioning from traditional MATLAB methods to Python's
advanced computational tools, such as automatic differentiation. We demonstrate Python's streamlined approach to minimizing nonlinear energies by analyzing three problem benchmarks - the p-Laplacian, the Ginzburg-Landau model, and the Neo-Hookean hyperelasticity. This approach merely requires the provision of the energy functional itself, making it a simple and efficient way to solve this category of problems. The results show that the implementation is about ten times faster than the MATLAB implementation for large-scale problems. Our findings highlight Python's
efficiency and ease of use in scientific computing, establishing it as a preferable choice for implementing sophisticated mathematical models and accelerating the development of numerical simulations.
\end{abstract}


\section{Introduction}
Solving problems posed by partial differential equations can often
be accomplished using the variational approach, which is based on
finding a minimum of the corresponding energy functional 
\begin{equation}
J(u)=\min_{v\in V}J(v)\,,\label{minim}
\end{equation}
where $V$ is a space of test functions defined in a domain $\Omega$
and includes Dirichlet boundary conditions on $\partial\Omega$. Problems
of this nature appear in various applications in physics and are mathematically
studied within the calculus of variations. The energy functionals
are then described by integrals over domains in two- or three-dimensional
space. The finite element method \cite{Ciarlet-FEM} can be utilized
as an approximation of (\ref{minim}) and results in a minimization
problem 
\begin{equation}
J(u_{h})=\min_{v\in V_{h}}J(v)\label{minim_discrete}
\end{equation}
formulated in the finite-dimensional subspace $V_{h}$ of $V.$ 

This contribution is based on recent MATLAB implementations \cite{MMV,MoVa,moskovka2022minimization},
which use the simplest linear nodal basis functions to enable an efficient solution of (\ref{minim_discrete}). 
It extends the implementation
to Python and exploits the new computational tools available therein,
namely automatic differentiation, efficient graph coloring, and algebraic
multigrid solvers. Our goal remains consistent with that of the MATLAB
implementation: to achieve efficient and vectorized computations without
the need for manual gradient implementation.

We reimplement and compare three problems: the p-Laplace problem,
the Ginzburg-Landau model in superconductivity \cite{Carstensen},
and the Neo-Hookean hyperelastic model in solid mechanics \cite{kruvzik2019mathematical}.

Solution times were obtained using an AMD Ryzen 9 7940HS processor
with 64 GB of memory, running MATLAB R2023b or Python 3.11.8 with packages (jax=0.4.26, numpy=1.26.4, scipy=1.13.0, igraph=0.11.4, pyamg=5.1.0).

The source code for the Python implementation is available on GitHub:
\begin{center}
\href{https://github.com/Beremi/nonlinear_energies_python/tree/PPAM2024}{https://github.com/Beremi/nonlinear\_energies\_python/tree/PPAM2024}
\end{center}

\section{Method description}
We provide an overview of the Python and MATLAB implementations,
highlighting their differences. Since both implementations are based
on the Finite Element Method (FEM) and use linear nodal elements, they share
many similarities and common requirements for some precomputed values
on the grid.

\begin{figure}
\vspace{-0.2cm}
\centering
\includegraphics[width=0.47\columnwidth]{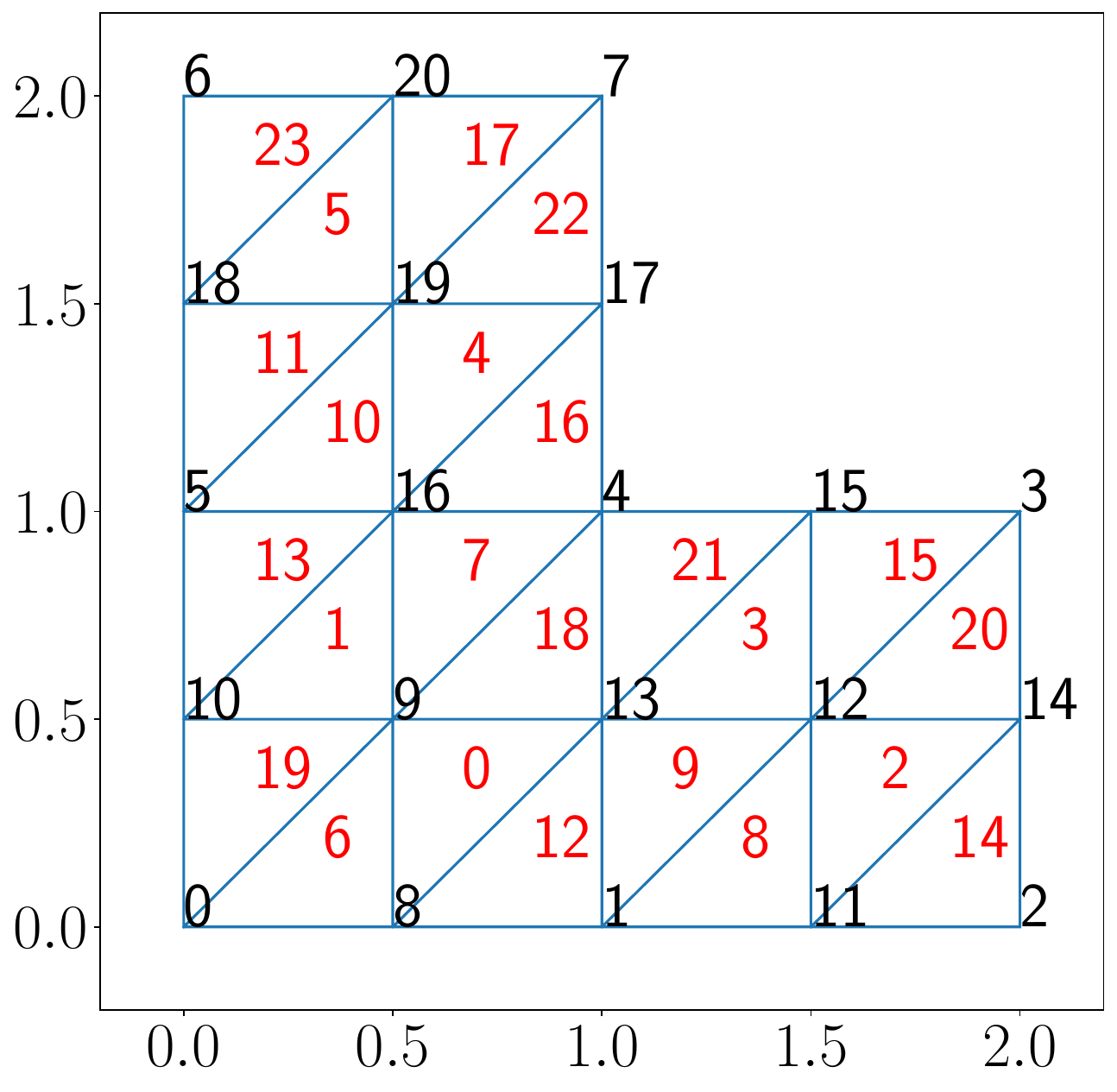}
\hfill
\includegraphics[width=0.47\columnwidth]{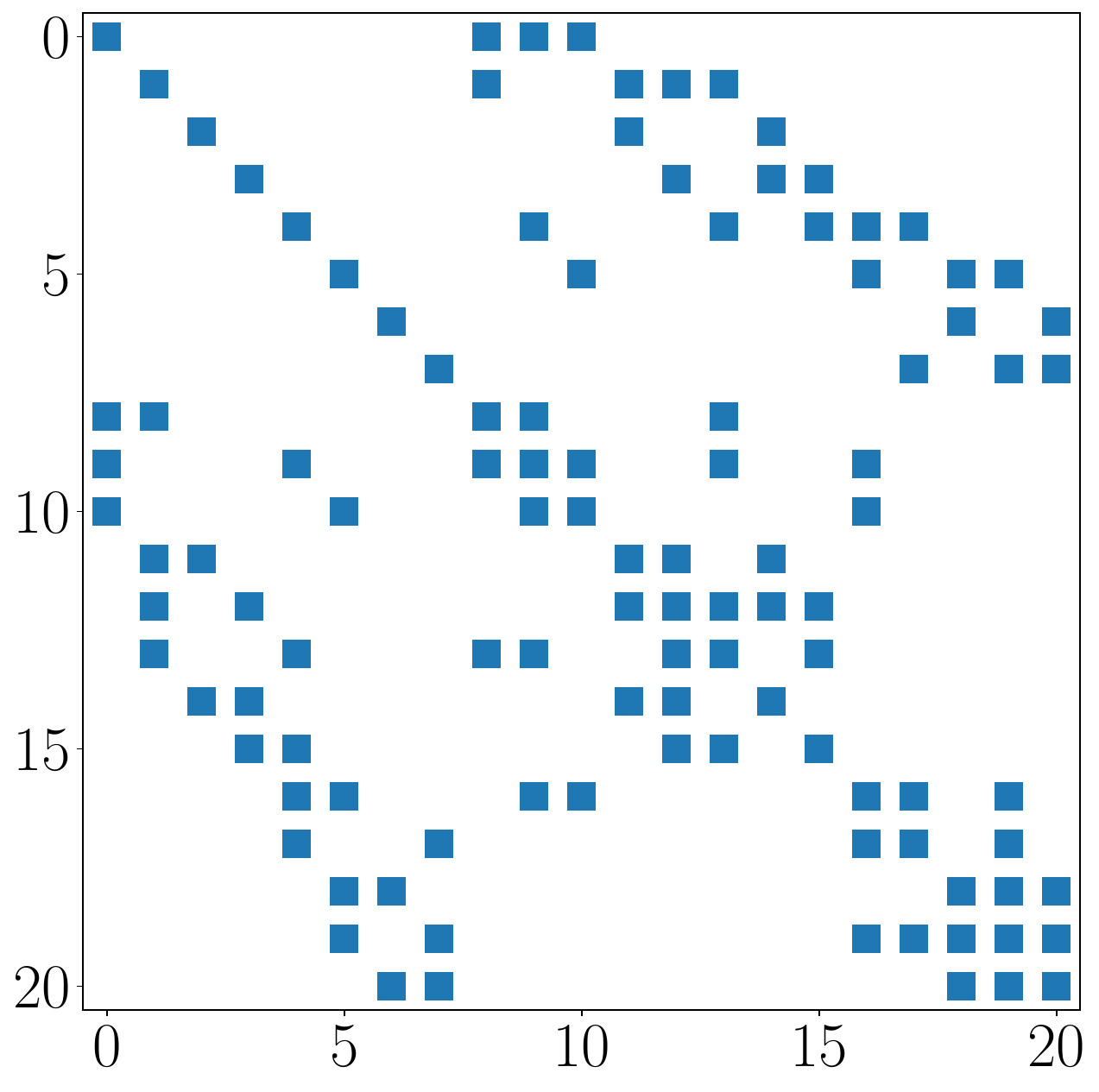}
\caption{Discretization of a rectangular domain (left) 
and the corresponding sparsity pattern (right).}
\label{fig:mesh_sparsity}
\vspace{-0.7cm}
\end{figure}

For a grid in 2D or 3D with $n$ - the number of elements, we will
use the following data as input to our energy functional implementations:
\begin{description}
\item [{\texttt{u:}}] Vector of minimized values for non-fixed degrees
of freedom. 
\item [{\texttt{u\_0:}}] Vector containing Dirichlet boundary values and
zeros for non-fixed degrees of freedom. 
\item [{\texttt{freedofs:}}] Vector of indices for nonfixed degrees of
freedom. 
\item [{\texttt{elems:}}] Matrix of indices for the corresponding nodes for
each element (triangle in 2D - $n\times3$, tetrahedron in 3D - $n\times4$). 
\item [{\texttt{dvx/dvy/dvz:}}] Matrices of the corresponding partial derivatives
$\frac{\partial v}{\partial x},\frac{\partial v}{\partial y},\frac{\partial v}{\partial z}$ of a test function
for each degree of freedom in each element (triangle in 2D - $n\times3$,
tetrahedron in 3D - $n\times4$). 
\item [{\texttt{vol:}}] Vector of areas or volumes of each element. 
\end{description}
Outside of these inputs to the energy functional implementation, there
is also a sparsity pattern incidence matrix computed specifically
from the grid. The sparsity pattern describes connections between
vertices (degrees of freedom) based on their support on elements;
i.e. a zero occurs only if there are no elements where both vertices
have support. An example of a coarse 2D triangular mesh together with its sparsity pattern is shown in Fig. \ref{fig:mesh_sparsity}.  It has 24 elements and 21 nodes. We aim for minimizations with up to 1 milion nodes on finer meshes. The sparsity pattern clearly corresponds to the sparsity pattern of Hessian $\nabla^2 J(v)$ for scalar problems discretized by linear nodal functions. 

\subsection{Overview of Matlab approach}
We briefly outline the main properties of the
current MATLAB implementation. 
\begin{itemize}
\item The gradient $\nabla J(v)$ is computed either analytically or via central differences approximation using energy densities
\cite{MoVa}. 
\item The minimization is fully performed by the MATLAB Optimization Toolbox
in the form of the \texttt{\textbf{fminunc}} function:
\begin{itemize}
\item It accepts the energy functional $J(v)$ and its gradient $\nabla J(v)$ along with the Hessian
sparsity pattern.
\item It uses SFD (Sparse Hessian via Finite Gradient Differences \cite{jacobian_color})
for the finite difference approximation of the Hessian $\nabla^2 J(v)$. 
\begin{itemize}
\item The coloring of the sparsity pattern needed for SFD is done using greedy
graph coloring and needs to be recomputed with each \texttt{\textbf{fminunc}}
function call. Additionally, the MATLAB implementation is very slow
for larger meshes. 
\end{itemize}
\item It employs a Trust Region with diagonally preconditioned CG (Conjugate
Gradient) solver, which is very inefficient for badly conditioned
matrices and can scale poorly with matrices whose conditional number
deteriorates with increasing grid size, which is typical for FEM. 
\end{itemize}
\end{itemize}

\subsection{Python approach and implementation}
We present the Python implementation, highlighting
mainly its differences from MATLAB's. We start by stating that the
energy functional is implemented almost identically. 
The Python implementations will be showcased in subsequent
sections. We used the JAX library with 64-bit floats. JAX is a Python library for accelerator-oriented array computation and program transformation, designed for high-performance numerical computing and large-scale machine learning.
Our Python implementation can be summarized as follows: 
\begin{itemize}
\item The gradient is computed via the JAX library \cite{jax}. This is
an automatic process; no additional code is needed, as everything
is handled by the \texttt{\textbf{jax.grad}} function. Note that the
gradient computed this way is the exact analytical gradient. 
\item The Hessian is computed similarly to the SFD \cite{jacobian_color}
approach with one major difference: Instead of using finite differences,
the exact Jacobian dot product is used in the form of the call to the \texttt{\textbf{jax.jvp}}
function. Note that the Hessian computed in this way is the exact
analytical Hessian.
\item Graph coloring is performed by the efficient package iGraph \cite{igraph}.
As this is our implementation of the minimizer procedure, the coloring
can be reused if the same grid is used to solve multiple problems. 
\item All costly operators, such as energy, its gradient and Hessian evaluations,
are compiled into XLA (Accelerated Linear Algebra) using \texttt{\textbf{jax.jit}},
resulting in faster execution. 
\begin{itemize}
\item This operation takes approximately 1.3 seconds across tested problems
and grids and is a one-time operation; there is no need to recompute
it for different grids. 
\end{itemize}
\item An energy minimization is performed using a basic textbook implementation of
the Newton method with line search using the Golden section method;
see \cite{newton_linesearch}. 
\item The solution of linear systems is accomplished using the Algebraic Multigrid
solver (AMG), employing the PyAMG package \cite{pyamg}, or using SciPy's direct solver. The choice of solver is determined by the size of the sparse system; systems with up to $15,000$ unknowns (degrees of freedom) are solved directly, whereas larger systems are addressed using AMG.

\end{itemize}

\section{Model problems and results}
We present three benchmark problems, showcase their implementation in Python, and compare the complexity of the solutions
to the current MATLAB implementation. 
For the Python implementation, we measure the setup time only once per mesh, which can be reused for multiple problems with different settings. This is not possible in the Matlab implementation.
The expectation for the performance of all problems is as follows: 
\begin{itemize}
\item automatically derived and compiled Jacobians and Hessians will be
slightly faster in evaluation than their MATLAB counterparts,
\item analytical Jacobians and Hessians can accelerate the convergence of
Newton minimization (or trust region) methods compared to energy density
and SFD approximations,
\item algebraic multigrid solvers will be faster and, for increasingly worse
conditioned problems with increasing grid size will have better scaling (compared to diagonally preconditioned CG).
\end{itemize}

\subsection{p-Laplace 2D}

The first benchmark problem is a (weak) solution of the p-Laplace
equation \cite{Lindqvist}: 
\begin{equation}
\begin{split}\Delta_{p}u & =f\quad\mbox{in}\:\Omega\,,\\
u & =g\quad\mbox{on}\:\partial\Omega,
\end{split}
\label{pLapl}
\end{equation}
where the p-Laplace operator is defined as 
\[
\Delta_{p}u=\nabla\cdot\left(\|\nabla v\|^{p-2}\nabla u\right)
\]
for some power $p>1$. 
The domain $\Omega\in\mathbb{R}^{d}$ is assumed
to have a Lipschitz boundary $\partial\Omega$, with $f\in L^{2}(\Omega)$
and $g\in W^{1-1/p,p}(\partial\Omega)$, where $L$ and $W$ denote
the standard Lebesgue and Sobolev spaces, respectively. It is known
that (\ref{pLapl}) represents an Euler-Lagrange equation corresponding
to a minimization problem 
\begin{equation}
J(u)=\min_{v\in V}J(v),\quad J(v):=\frac{1}{p}\int\limits_{\Omega}\|\nabla v\|^{p} \dx-\int\limits_{\Omega}f\,v \dx,\label{energy_pLaplace}
\end{equation}
where $V=W_{g}^{1,p}(\Omega)=\{v\in W^{1,p},v=g\mbox{ on }\partial\Omega\}$
includes Dirichlet boundary conditions on $\partial\Omega$. The minimizer
$u\in V$ of \eqref{energy_pLaplace} is known to be unique for $p>1$.

\begin{figure}[ht]
\vspace{-0.2cm}
\begin{centering}
\includegraphics[width=0.65\columnwidth]{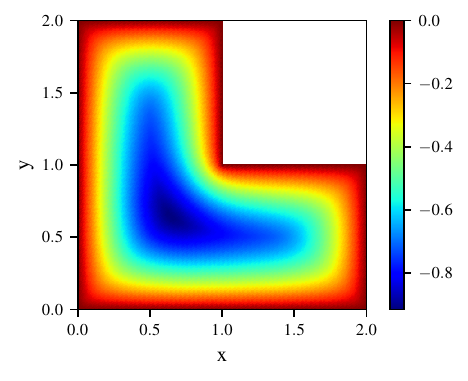}
\par\end{centering}
\caption{Numerical approximation of the solution $u$ of the p-Laplace benchmark.}\label{fig:pLaplace_sol}
\vspace{-0.7cm}
\end{figure}

\begin{figure}[H]
\begin{centering}
\begin{tabular}{|r||c|r|r||r|r|r|}
\cline{2-7}
\multicolumn{1}{r|}{} & \multicolumn{3}{c||}{\textbf{Python}} & \multicolumn{3}{c|}{\textbf{Matlab}}\tabularnewline
\hline 
\multicolumn{1}{|r|}{\textbf{dofs}} & \textbf{time {[}s{]}} & \textbf{iters} & \textbf{$J\left(\boldsymbol{u}\right)$} & \textbf{time {[}s{]}} & \textbf{iters} & \textbf{$J\left(\boldsymbol{u}\right)$}\tabularnewline
\hline 
\hline 
$33$ & 0.01 / 0.18 & 4 & -7.3411 & 0.05 & 8 & -7.3411\tabularnewline
\hline 
$161$ & 0.01 / 0.17 & 4 & -7.7767 & 0.09 & 10 & -7.7767\tabularnewline
\hline 
$705$ & 0.02 / 0.17 & 5 & -7.9051 & 0.16 & 11 & -7.9051\tabularnewline
\hline 
$2,945$ & 0.07 / 0.25 & 6 & -7.9430 & 0.49 & 11 & -7.9430\tabularnewline
\hline 
$12,033$ & 0.30 / 0.23 & 6 & -7.9546 & 1.47 & 11 & -7.9546\tabularnewline
\hline 
$48,641$ & 0.86 / 0.42 & 6 & -7.9583 & 5.12 & 11 & -7.9583\tabularnewline
\hline 
$195,585$ & 4.14 / 1.20 & 8 & -7.9596 & 41.90 & 11 & -7.9596\tabularnewline
\hline 
$784,385$ & 19.96 / 4.41 & 9 & -7.9600 & 429.12 & 13 & -7.9600\tabularnewline
\hline 
\end{tabular}
\par\end{centering}
\captionof{table}{Performance comparison 
for the p-Laplace benchmark}\label{tab:plaplace}

\begin{centering}
\includegraphics[bb=20bp 0bp 275bp 181bp]{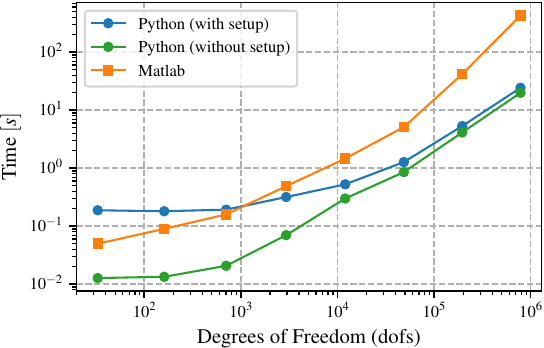}
\par\end{centering}
\captionof{figure}{Performance comparison 
for the p-Laplace benchmark}\label{fig:plaplace}
\vspace{-0.7cm}
\end{figure}

Assume $f=-10$ for $x\in\Omega$, where $\Omega=\left\langle 0,2\right\rangle ^{2}\setminus\left\langle 1,2\right\rangle ^{2}\subset\mathbb{R}^{2}$ is a L-shaped domain,
$p=3$, and homogeneous Dirichlet boundary conditions on the boundary $\partial\Omega$.
The exact solution $u$ of \eqref{energy_pLaplace} is unknown but can be
approximated numerically; see Figure \ref{fig:pLaplace_sol}. The FEM approximation for this setting serves
as a computational benchmark. 

The complete implementation of the energy operator of \eqref{energy_pLaplace} in JAX can be seen
in the Listing \ref{lis:plaplace}. The corresponding gradient and Hessian are then evaluated exactly and automatically. The additional
parameters are the power $\mathrm{p}=p$ and the vector $\mathrm{f}=\left[f_{1},\ldots,f_{n}\right]$,
where $f_{i}=\int\limits_{\Omega}f\,v_{i} \dx$ with each $v_{i}\in V_{h}$. The vector $\mathrm{f}$ represents the linear functional $f$ in terms of the FEM basis $v_{i}\in V_{h}$.

The resulting times and their comparison to the original MATLAB implementation
can be seen in Table \ref{tab:plaplace} and Figure \ref{fig:plaplace}.
The observations align with the expectation that the Python implementation
is faster, albeit with a fixed cost incurred by assembling and compiling
functions for gradient and Hessian evaluation. 

\begin{lstlisting}[caption={Evaluation of the p-Laplace energy},label={lis:plaplace},language=Python,float=htb,numbers=left,basicstyle={\small\ttfamily}]
def J(u, u_0, freedofs, elems, dvx, dvy, vol, p, f):
    v = u_0.at[freedofs].set(u)
    v_elems = v[elems]

    F_x = jnp.sum(v_elems * dvx, axis=1)
    F_y = jnp.sum(v_elems * dvy, axis=1)

    intgrds = (1 / p) * (F_x**2 + F_y**2)**(p / 2)
    return jnp.sum(intgrds * vol) - jnp.dot(f, v)
\end{lstlisting}

\subsection{Ginzburg-Landau problem}

We consider the Ginzburg-Landau minimization problem \cite{Carstensen}
for a scalar test function $v\in V$. The energy functional is given
by 
\begin{equation}
J(v)=\intop_{\Omega}\left(\frac{\varepsilon}{2}\|\nabla v\|^{2}+\frac{1}{4}(v^{2}-1)^{2}\right) \dx,\label{energy_GL}
\end{equation}
where $\Omega\subset\mathbb{R}^{d}$ is a given domain and $\varepsilon$
is a given small positive parameter.

The benchmark problem consists of $\Omega=\left\langle -1,1\right\rangle ^{2}\subset\mathbb{R}^{2}$
and $\varepsilon=0.01$. The Python implementation that allows for
automatic differentiation of the energy functional \eqref{energy_GL} can be found in Listing \ref{lis:GL}. Additional
non-grid parameters are 
\[
\renewcommand{\arraystretch}{1.2}
\mathrm{ip}=\left[\begin{array}{ccc}
\frac{2}{3} & \frac{1}{6} & \frac{1}{6}\\
\frac{1}{6} & \frac{2}{3} & \frac{1}{6}\\
\frac{1}{6} & \frac{1}{6} & \frac{2}{3}
\end{array}\right],
\quad 
\mathrm{w}=\left[\begin{array}{c}
\frac{1}{3}\\
\frac{1}{3}\\
\frac{1}{3}
\end{array}\right],
\quad \mathrm{eps}=\varepsilon,
\]
where $\mathrm{ip}$ is the matrix of coefficient of integration points and $\mathrm{w}$ is the vector of their weigths. The integration rule used here is the same as in the corresponding MATLAB code. Note that it is accurate only up to the second degree of the polynomial, but $(v^{2}-1)^{2}$ requires the exact integration of fourth-order polynomials. This "variational crime" is acceptable, as linear convergence is expected with the linear elements considered.

\begin{figure}[H]
\begin{centering}
\includegraphics[width=0.65\columnwidth]{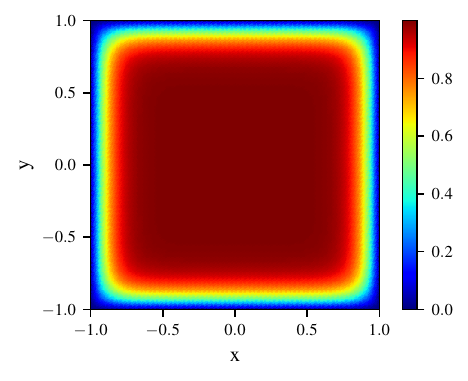}
\par\end{centering}
\caption{Numerical approximation of the solution $u$ of the G-L benchmark.}\label{sol:GL}
\end{figure}

\vspace{-1cm}
\begin{lstlisting}[caption={Evaluation of the Ginzburg-Landau energy},label={lis:GL},language=Python,float=htb,numbers=left,basicstyle={\small\ttfamily}]
def J(u, u_0, freedofs, elems, dvx, dvy, vol, ip, w, eps):
    v = u_0.at[freedofs].set(u)
    v_elems = v[elems]
    
    F_x = jnp.sum(v_elems * dvx, axis=1)
    F_y = jnp.sum(v_elems * dvy, axis=1)

    e_1 = (1 / 2) * eps * (F_x**2 + F_y**2)
    e_2 = (1 / 4) * ((v_elems @ ip)**2 - 1)**2 @ w
    return jnp.sum((e_1 + e_2) * vol)
\end{lstlisting}

\begin{figure}[ht]
\begin{centering}
\begin{tabular}{|r||c|r|r||r|r|r|}
\cline{2-7}
\multicolumn{1}{r|}{} & \multicolumn{3}{c||}{\textbf{Python}} & \multicolumn{3}{c|}{\textbf{Matlab}}\tabularnewline
\hline 
\multicolumn{1}{|r|}{\textbf{dofs}} & \textbf{time {[}s{]}} & \textbf{iters} & \textbf{$J\left(\boldsymbol{u}\right)$} & \textbf{time {[}s{]}} & \textbf{iters} & \textbf{$J\left(\boldsymbol{u}\right)$}\tabularnewline
\hline 
\hline 
\textbf{$49$} & 0.01 / 0.18 & 6 & 0.3867 & 0.07  & 8 & 0.3867\tabularnewline
\hline 
\textbf{$225$} & 0.02 / 0.18 & 8 & 0.3547 & 0.04  & 6 & 0.3547\tabularnewline
\hline 
\textbf{$961$} & 0.03 / 0.24 & 7 & 0.3480 & 0.13  & 7 & 0.3480\tabularnewline
\hline 
\textbf{$3,969$} & 0.10 / 0.20 & 7 & 0.3462 & 0.27  & 6 & 0.3462\tabularnewline
\hline 
\textbf{$16,129$} & 0.40 / 0.26 & 6 & 0.3458 & 0.60 & 7 & 0.3458\tabularnewline
\hline 
\textbf{$65,025$} & 1.47 / 0.51 & 6 & 0.3457 & 4.65  & 8 & 0.3457\tabularnewline
\hline 
\textbf{$261,121$} & 6.04 / 1.56 & 6 & 0.3456 & 64.68  & 8 & 0.3456\tabularnewline
\hline 
\textbf{$1,046,529$} & 26.63 / 5.76 & 6 & 0.3456 & 653.08  & 9 & 0.3456 \tabularnewline
\hline 
\end{tabular}
\par\end{centering}
\captionof{table}{Performance comparison 
for the G-L benchmark}\label{tab:GL}

\begin{centering}
\includegraphics[bb=20bp 0bp 274bp 182bp]{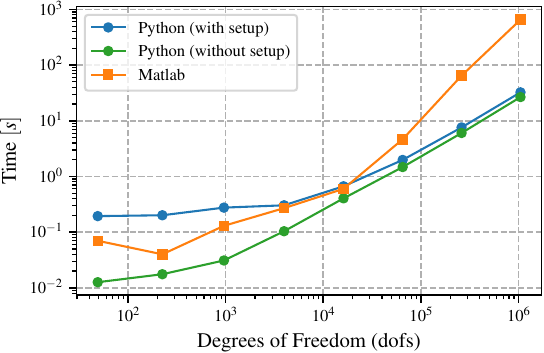}
\par\end{centering}
\captionof{figure}{Performance comparison 
for the G-L benchmark}\label{fig:GL}
\end{figure}

The resulting comparison with the Matlab implementation is presented in Table \ref{tab:GL} and Figure \ref{fig:GL}. We observed faster solution times; however, this was the only scenario in which the number of iterations for the analytical gradient and Hessian matched those of their numerical approximations. This behavior was primarily attributed to the choice of methods, as the diagonally regularized matrices in the trust-region approach yielded improved descent directions. Despite this, the differences in performance were minimal and the computation times were nearly identical for both the line-search and trust-region methods. Consequently, we opted to present only one method in the Python implementation.

\subsection{Hyperelasticity in 3D}

The last benchmark is based on minimization of the energies of
hyperelastic materials in mechanics of solids \cite{kruvzik2019mathematical}.
The trial space is chosen as $V=W_{D}^{1,p}(\Omega,\mathbb{R}^{\text{dim}})$,
that is, the (vector) Sobolev space of $L^{p}$ integrable functions
with the first weak derivative also being $L^{p}$ integrable and
satisfying (in the sense of traces) Dirichlet boundary conditions
$v(x)=u_{p}(x)$ at the domain boundary $x\in\partial\Omega$, for
a prescribed function $u_{p}:\partial\Omega\to\mathbb{R}^{\text{dim}}$.
A primary variable is the deformation mapping $v\in V$, which describes
the relocation of any point $x\in\Omega$ during the deformation process.

The gradient deformation tensor $F\in L^{p}(\Omega,\mathbb{R}^{\text{dim}\times\text{dim}})$
is defined as 
\begin{equation}
F(v)=\nabla v=\begin{bmatrix}\frac{\partial v^{(1)}}{\partial x_{1}} & \cdots & \frac{\partial v^{(1)}}{\partial x_{\text{dim}}}\\
\vdots & \ddots & \vdots\\
\frac{\partial v^{(\text{dim})}}{\partial x_{1}} & \cdots & \frac{\partial v^{(\text{dim})}}{\partial x_{\text{dim}}}
\end{bmatrix}
\end{equation}
The energy functional is given by 
\begin{align}
J\left(v\right) & =\int_{\Omega}W(F(v(x))) \dx-\int_{\Omega}f(x)\cdot v(x) \dx,
\end{align}
where $W:\mathbb{R}^{\text{dim}\times\text{dim}}\to\mathbb{R}$ defines
a strain-energy density function, and $f:\Omega\to\mathbb{R}^{\text{dim}}$
is a loading functional. We assume the compressible Neo-Hookean density
\begin{equation}
W(F)=C_{1}(I_{1}(F)-\text{dim}-2\log(\det F))+D_{1}(\det F-1)^{2},
\end{equation}
where $I_{1}(F)=|F|^{2}$ uses the Frobenius norm $|\cdot|$, and
$\det(\cdot)$ is the matrix determinant operator. 

As a benchmark in $\text{dim}=3$, we consider a bar domain 
$$\Omega=(0,l_{x})\times\left(-\frac{l_{y}}{2},\frac{l_{y}}{2}\right)\times\left(-\frac{l_{z}}{2},\frac{l_{z}}{2}\right), \qquad \text{where }l_{x}=0.4, l_{y}=0.01,
$$
defined by the equivalent pairs of material parameters
\begin{itemize}
    \item $E=2\cdot10^{8}$ (Young's modulus), $\nu=0.3$ (Poisson's ratio),
    \item $\mu=\frac{E}{2(1+\nu)}$ (the shear modulus), $K=\frac{E}{3(1-2\nu)}$ (the bulk modulus),
    \item $C_{1}=\frac{\mu}{2},D_{1}=\frac{K}{2}$.
\end{itemize}
No loading is assumed, $f=0$. 
The bar undergoes a deformation characterized by prescribed deformations
at the right end of the bar, involving clockwise rotations up to 4
full turns across 24 iterations. The left end of the bar remains intact. 

\begin{figure}[H]
\begin{centering}
\includegraphics[width=0.98\columnwidth]{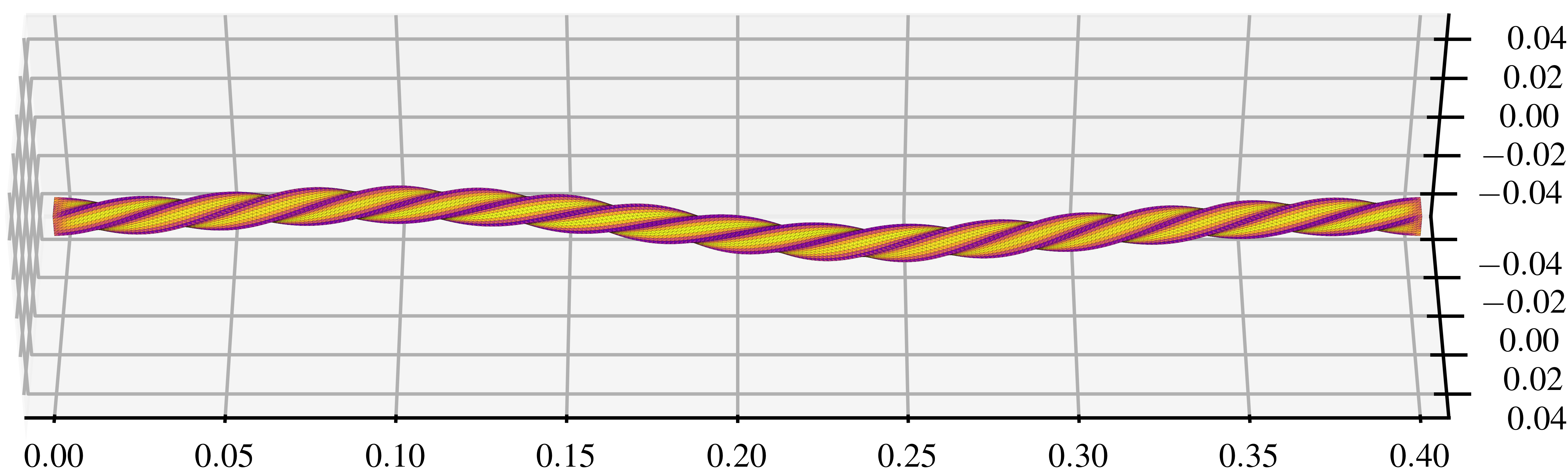}
\par\end{centering}
\caption{Solution of hyperelasticity benchmark with underlying Neo-Hook densities}\label{fig:Hyper_sol}
\end{figure}

The Python implementation that facilitates automatic differentiation
can be found in Listing \ref{lis:hyper}. Additional non-grid parameters
are $\mathrm{C1}=C_{1}$ and $\mathrm{D1}=D_{1}$.

\begin{lstlisting}[caption={Evaluation of the hyperelastic energy},label={lis:hyper},language=Python,float=htb,numbers=left,basicstyle={\small\ttfamily}]
def J(u, u_0, freedofs, elems, dvx, dvy, dvz, vol, C1, D1):
    v = u_0.at[freedofs].set(u)
    vx_elem = v[0::3][elems]
    vy_elem = v[1::3][elems]
    vz_elem = v[2::3][elems]

    F11 = jnp.sum(vx_elem * dvx, axis=1)
    F12 = jnp.sum(vx_elem * dvy, axis=1)
    F13 = jnp.sum(vx_elem * dvz, axis=1)
    F21 = jnp.sum(vy_elem * dvx, axis=1)
    F22 = jnp.sum(vy_elem * dvy, axis=1)
    F23 = jnp.sum(vy_elem * dvz, axis=1)
    F31 = jnp.sum(vz_elem * dvx, axis=1)
    F32 = jnp.sum(vz_elem * dvy, axis=1)
    F33 = jnp.sum(vz_elem * dvz, axis=1)

    I1 = (F11**2 + F12**2 + F13**2 + 
          F21**2 + F22**2 + F23**2 + 
          F31**2 + F32**2 + F33**2)
    det = jnp.abs(+ F11 * F22 * F33 - F11 * F23 * F32
                  - F12 * F21 * F33 + F12 * F23 * F31
                  + F13 * F21 * F32 - F13 * F22 * F31)
    W = C1 * (I1 - 3 - 2 * jnp.log(det)) + D1 * (det - 1)**2
    return jnp.sum(W * vol)
\end{lstlisting}

\begin{figure}
\begin{centering}
\begin{tabular}{|c|c||r|r|r||r|r|r|}
\cline{3-8}
\multicolumn{1}{c}{} & \multicolumn{1}{c|}{} & \multicolumn{3}{c||}{\textbf{Python}} & \multicolumn{3}{c|}{\textbf{Matlab}}\tabularnewline
\hline 
\textbf{level} & \multicolumn{1}{c|}{$t$} & \textbf{time {[}s{]}} & \textbf{iters} & \textbf{$J\left(\boldsymbol{u}\right)$} & \textbf{time {[}s{]}} & \textbf{iters} & \textbf{$J\left(\boldsymbol{u}\right)$}\tabularnewline
\hline 
\hline 
\multirow{8}{*}{\begin{cellvarwidth}[t]
\centering
\textbf{1:}\\
\textbf{$2,133$}\\
\textbf{dofs}
\end{cellvarwidth}} & 3 & 0.25 & 19 & 3.1173 & 6.47 & 51 & 3.1173\tabularnewline
\cline{2-8}
 & 6 & 0.26 & 20 & 12.4423 & 8.38 & 61 & 12.4424\tabularnewline
\cline{2-8}
 & 9 & 0.27 & 21 & 27.8990 & 9.49 & 75 & 27.8992\tabularnewline
\cline{2-8}
 & 12 & 0.26 & 20 & 49.5501 & 10.51 & 85 & 49.5508\tabularnewline
\cline{2-8}
 & 15 & 0.30 & 23 & 77.3831 & 10.39 & 81 & 77.3862\tabularnewline
\cline{2-8}
 & 18 & 0.28 & 21 & 111.3262 & 11.03 & 89 & 111.3353\tabularnewline
\cline{2-8}
 & 21 & 0.30 & 23 & 151.4552 & 9.50 & 78 & 151.4901\tabularnewline
\cline{2-8}
 & 24 & 0.62 & 47 & 197.7484 & 9.89 & 81 & 197.7606\tabularnewline
\hline 
\hline 
\multirow{8}{*}{\begin{cellvarwidth}[t]
\centering
\textbf{2:}\\
\textbf{$11,925$}\\
\textbf{dofs}
\end{cellvarwidth}} & 3 & 4.55 & 23 & 1.8244 & 44.08 & 47 & 1.8244\tabularnewline
\cline{2-8}
 & 6 & 4.70 & 24 & 7.2960 & 61.72 & 64 & 7.2961\tabularnewline
\cline{2-8}
 & 9 & 4.36 & 23 & 16.4069 & 59.08 & 62 & 16.4088\tabularnewline
\cline{2-8}
 & 12 & 4.00 & 21 & 29.1607 & 96.18 & 98 & 29.1673\tabularnewline
\cline{2-8}
 & 15 & 4.09 & 21 & 45.5598 & 72.29 & 73 & 45.5696\tabularnewline
\cline{2-8}
 & 18 & 4.29 & 22 & 65.5437 & 64.18 & 68 & 65.5493\tabularnewline
\cline{2-8}
 & 21 & 4.53 & 23 & 89.1459 & 55.33 & 57 & 89.1683\tabularnewline
\cline{2-8}
 & 24 & 7.57 & 39 & 116.3232 & 56.91 & 64 & 116.3278\tabularnewline
\hline 
\hline 
\multirow{8}{*}{\begin{cellvarwidth}[t]
\centering
\textbf{3 :}\\
\textbf{$77,517$}\\
\textbf{dofs}
\end{cellvarwidth}} & 3 & 75.38 & 24 & 1.4631 & 861.14 & 58 & 1.4631\tabularnewline
\cline{2-8}
 & 6 & 77.67 & 24 & 5.8533 & 1008.09 & 67 & 5.8533\tabularnewline
\cline{2-8}
 & 9 & 73.92 & 23 & 13.1731 & 941.76 & 65 & 13.1734\tabularnewline
\cline{2-8}
 & 12 & 69.71 & 21 & 23.4252 & 915.22 & 63 & 23.4284\tabularnewline
\cline{2-8}
 & 15 & 74.70 & 23 & 36.6100 & 1024.90 & 71 & 36.6235\tabularnewline
\cline{2-8}
 & 18 & 70.81 & 22 & 52.7169 & 1079.76 & 75 & 52.7635\tabularnewline
\cline{2-8}
 & 21 & 98.70 & 29 & 71.7516 & 1278.45 & 88 & 71.7834\tabularnewline
\cline{2-8}
 & 24 & 171.82 & 49 & 93.7039 & 1837.23 & 128 & 93.7216\tabularnewline
\hline 
\end{tabular}
\par\end{centering}
\captionof{table}{Performance comparison for the hyperelasticity benchmark}\label{tab:hyper}


\begin{centering}
\includegraphics[bb=20bp 0bp 275bp 180bp]{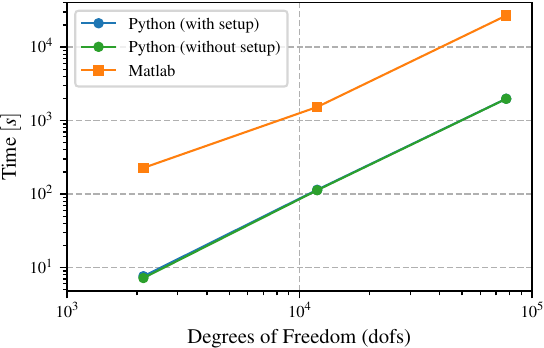}
\par\end{centering}
\captionof{figure}{Performance comparison for the hyperelasticity benchmark}\label{fig:hyper}
\end{figure}

The comparative analysis with the MATLAB implementation is detailed in Table \ref{tab:hyper} and Figure \ref{fig:hyper}. Additionally, the Python setup times are as follows:

\begin{itemize}
    \item $0.42$ s for \textbf{$2,133$} degrees of freedom (dofs),
    \item $1.69$ s for \textbf{$11,925$} dofs, and
    \item $4.86$ s for \textbf{$77,517$} dofs.
\end{itemize}
These setup durations are almost negligible for this particular problem, given that the solution process---which encompasses all 24 iterations---is significantly more computationally demanding by comparison. The speedups are particularly notable for this benchmark, ranging from around $10\times$ to $20\times$ faster than the MATLAB computations. It is critical to note that MATLAB failed to converge with the same precision as the Python code. For comparison purposes, we maintain the original settings of the MATLAB code. Achieving the same level of convergence could be possible by increasing the tolerance; however, this adjustment would substantially extend the computation times.

\section{Conclusions}
We have presented a comparison of the MATLAB and Python
implementations to solve a class of complex minimization problems.
Our findings highlight several key advantages of using Python in scientific computing:

Firstly, the Python framework facilitates the implementation of new
nonlinear functionals with remarkable ease, often requiring only a
few lines of code. This simplicity significantly enhances the flexibility and speed of development in research settings.

Second, we have identified that custom solvers for linear systems,
particularly those involving Hessians, are highly preferable. Unfortunately,
MATLAB’s \texttt{\textbf{fminunc}} function, with its trust region
approach that uses only a sparse pattern and diagonally preconditioned
conjugate gradients, does not support custom linear solvers. This limitation would require MATLAB implementations to rewrite the minimizer routine
to their own implementations that facilitate these options. However,
there are still no easily obtainable and efficient implementations
for things such as algebraic multigrid solvers or efficient graph
coloring algorithms.

Lastly, our benchmarks demonstrate that the current Python
framework is more than ten times faster than MATLAB for larger problems.

These observations suggest that the current Python environment provides a more flexible and efficient computational environment for handling sophisticated mathematical models and large-scale problems.

\bibliographystyle{plain}
\bibliography{references}
\end{document}